\documentclass[showpacs,preprintnumbers,twocolumn,amsmath,amssymb,prl]{revtex4}

\usepackage{epsfig}

\begin{document}

\title{Density Matrix Renormalization Group study of $^{48}$Cr and $^{56}$Ni}

\author{B. Thakur\dag\, S. Pittel\dag\  and N.Sandulescu\ddag\ }

\address{\dag\ Bartol Research Institute and Department of Physics and Astronomy, University of Delaware,
Newark, DE 19716, USA}

\address{\ddag\ Institute of Physics and Nuclear Engineering,
76900, Bucharest, Romania}

\date{\today}

%%%%%%%%%%%%%%%%%%
%%   ABSTRACT   %%
%%%%%%%%%%%%%%%%%%s

\begin{abstract}

We discuss the development of an angular-momentum-conserving
variant of the Density Matrix Renormalization Group (DMRG) method
for use in large-scale shell-model calculations of atomic nuclei
and report a first application of the method to the ground state
of $^{56}$Ni and improved results for $^{48}$Cr. In both cases, we
see a high level of agreement with the exact results. A comparison
of the two shows a dramatic reduction in the fraction of the space
required to achieve accuracy as the size of the problem grows.
\end{abstract}

 \maketitle

%%%%%%%%%%%%%%%%%%
%% INTRODUCTION %%
%%%%%%%%%%%%%%%%%%
% \section{Introduction}

One of the foremost challenges confronting nuclear physics today
is the systematic study of medium-mass and heavy nuclei using the
shell model. Even with the drastic truncation achieved by limiting
the active particles to one or at most a couple of valence shells
outside a doubly magic core, the size of the resulting space still
exceeds storage capabilities of the currently available
computational resources for all but fairly light nuclei. This
opens up the need for innovative truncation strategies. The
Density Matrix Renormalization Group (DMRG) method has had
outstanding success dealing with low-dimensional quantum lattice
problems \cite{Whi92}. The method was later extended to finite
fermi systems, where it has been applied with impressive success
to the description of small metallic grains \cite{Duk99}, to
two-dimensional electrons in strong magnetic fields \cite{Shib01},
and to problems in quantum chemistry \cite{Whi99}. This suggests
that it might also prove useful in the description of another
finite fermi system, the atomic nucleus. In this work we present
our first results for $^{56}$Ni, the most ambitious test we have
considered to date, and confirm its usefulness for nuclear
structure calculations.

%%%%%%%%%%%%%%%%%%%%%%%%%%
%% DMRG Concise details %%
%%%%%%%%%%%%%%%%%%%%%%%%%%

The usual DMRG algorithm begins by partitioning the complete
Hilbert space for a given problem onto a set of lattice sites,
with each site admitting a set of basis states. The untruncated
problem is recovered by considering the complete set of states
within all sites. Since the full space is typically too large for
exact treatment, the DMRG method treats the problem iteratively,
by successively adding sites to those already treated and
implementing a truncation based on density matrix considerations.
This is illustrated schematically if Figure 1.

% \begin{figure}
% \begin{tabular}{lllr}
% Sites               & $ \bullet_{L_1} \cdots \bullet_{L_r}  \bullet_{L_{r+1}}$ %
%                     & $\bullet_{L_{r+2}} \cdots \bullet_{L_n}$&\\
% 
% System Block $B_r$  & \fbox{$ \bullet_{L_1} \cdots \bullet_{{L_r}} $} & &\\ %
% 
% Enlarged Block $B_{r+1}$%
%                     & \fbox{$ \left[ \bullet_{L_1} \cdots \bullet_{{L_r}}  \right]  \bullet_{L_{r+1}} $} &  \\
% Medium              & & \fbox{$ \bullet_{L_{r+2}} \cdots \bullet_{L_n}$} & \\
% 
% Superblock                 & \fbox{$ \left[ \bullet_{L_1} \cdots \bullet_{{L_r}}  \right]  \bullet_{L_{r+1}} $} %
%                     & \fbox{$ \bullet_{L_{r+2}} \cdots \bullet_{L_n}$} \\
% 
% \end{tabular}
% \caption{Schematic illustration of the DMRG growth procedure.
% Shown are all the sites in the lattice chain, those of the system
% block, those added in the enlargement process, those of the
% medium, and their organization in the superblock.
% }\label{fig:Lattice}
% \end{figure}

\begin{figure}
\epsfxsize=.95\columnwidth\epsfbox{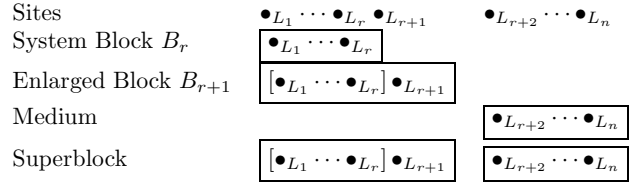}
\caption{Schematic illustration of the DMRG growth procedure.
Shown are all the sites in the lattice chain, those of the system
block, those added in the enlargement process, those of the
medium, and their organization in the superblock.
}\label{fig:Lattice}
\end{figure}

Assume that there are $n$ sites on the lattice and that $r$ sites
to the left have already been treated, defining what we call the
system block $B_r$. Assume further that within this block we have
$m$ states and the matrix elements of all suboperators of the
hamiltonian for these states. The next step would be to enlarge
$B_r$ by adding the $r+1^{st}$ site, producing the enlarged block
$ B_{r+1} \equiv B_r \otimes L_{r+1} $. We then couple this to a
medium, which involves information on all of the remaining sites,
producing the Superblock $B_r \otimes L_{r+1} \otimes M $,
representing the entire system.  The basic idea of the DMRG method
is to truncate the enlarged block to its optimum $m$ states,
namely those that are the most significant contributors to the
Superblock ground state, and to renormalize all operators to act
in this truncated space.

We will assume for now a product space description and denote the
states of the enlarged block as $|I\rangle_{B_{r+1}} =
|i\rangle_{B_r}|j\rangle_{r+1}$, where $i$ span the states of the
system block and $j$ those of the added site. The Superblock
ground state can then be written as
\begin{equation}
| \Psi_g \rangle_{SB} = \sum_{I,~k} B_{I,~k} |I \rangle_{B_{r+1}}
| k \rangle_M ~,
\end{equation}
where $k$ spans the states of the medium ($M$).

The reduced density matrix for the enlarged block in the
Superblock ground state is obtained by contracting over the states
of the medium, namely
\begin{equation}
\rho_{I,~I^\prime} = \sum_{k} {B^{*}_{I,~k}} B_{I^\prime,~k} ~.
\end{equation}
If we diagonalize this reduced density matrix and maintain the
eigenstates associated with its $m$ largest eigenvalues we are
guaranteed to have found the $m$ most important components of the
enlarged block in the Superblock ground state.

The basic idea then is to systematically grow the system block by
adding lattice sites and then at each stage to truncate to the $m$
most important states obtained in this way. At each stage we must,
as just noted, transform all hamiltonian suboperators to the
$m-$dimensional truncated space, as this provides required
information for its subsequent enlargement. The enlargement
process just described is repeated over and over, sweeping in both
directions through the set of lattice sites until convergence in
the ground-state energy is obtained. The calculations can then be
redone as a function of $m$ until acceptably small changes with
increasing values are obtained.

%%%%%%%%%%%%%%%%%%%%%%%%%%
%% JDMRG                %%
%%%%%%%%%%%%%%%%%%%%%%%%%%
% \section{J-DMRG}

The usual DMRG procedure, as just outlined, works with product
states and is thus equivalent to an m-scheme approach in the
nuclear context. This has the drawback that conserving angular
momentum symmetry becomes difficult. To avoid this problem in
applications to nuclei, Dukelsky and Pittel \cite{Pit04} proposed
the use of an angular-momentum-preserving variant of the DMRG,
called the JDMRG. This method, which is an example of a
non-Abelian DMRG algorithm \cite{McCulloch}, was first applied to
nuclei in the context of the Gamow Shell Model \cite{Naz04} and
then subsequently applied to the traditional nuclear shell model
by Pittel and Sandulescu \cite{Pit06}.

In the JDMRG, we work in a coupled (or J-scheme) basis. As a
consequence, we must now calculate reduced matrix elements of the
various suboperators of the Hamiltonian, namely
\begin{center}
$   a^\dagger_i, (a^\dagger_i a^\dagger_j )^\lambda, (a^\dagger_i \tilde{a}_j )^\lambda,  %
 {[(a^\dagger_i   a^\dagger_j )^\lambda \tilde{a}_k ]}^\kappa,$ \\%
$ [(a^\dagger_i   a^\dagger_j )^\lambda(\tilde{a}_k \tilde{a}_l)^\lambda ]^0 + h.c.$
\end{center}

An important feature of nuclei is that they contain two types of
particles, neutrons and protons. This leads to the question of how
the associated orbitals should be arranged on the lattice,
intertwined or at opposite ends.  We have found it useful to
maintain a separation between the neutron and proton blocks,
leading to what we call a three-block JDMRG algorithm, for reasons
to be made clear a bit later.

The nuclei for which we will present results here are $^{48}$Cr
and $^{56}$Ni. In both we assume a doubly magic $^{40}$Ca core
with the remaining nucleons distributed over the orbitals of the
$2p-1f$ shell.  We now spell out in a bit more detail the steps of
the JDMRG procedure that were implemented in our treatment of
these two nuclei.

\begin{flushleft}
{\it Setting the order of orbits:}
\end{flushleft}
As just noted, we have chosen to define the lattice chain such
that the neutron and the proton orbits lie on its opposite ends.
This still leaves several possible options, however. We have found
the optimal order to be the one shown in Figure 2, where we use
the notation $n_j$ and $p_j$ to denote a neutron orbital with
angular momentum $j$ and a proton orbital with angular momentum
$j$, respectively.

This order places maximally entangled orbits next to one another,
as is found to be important from quantum information
considerations \cite{Legeza}.

% \begin{figure}
% \begin{center}
% $\bullet_{\text{ n}_{\frac{7}{2}} } %
%  \bullet_{\text{ n}_{\frac{3}{2}} } %
%  \bullet_{\text{ n}_{\frac{1}{2}} } %
%  \bullet_{ \text{ n}_{\frac{5}{2}}}| %
%  \circ_{   \text{ p}_{\frac{5}{2}} }   %
%  \circ_{   \text{ p}_{\frac{1}{2}} }   %
%  \circ_{   \text{ p}_{\frac{3}{2}} }   %
%  \circ_{   \text{ p}_{\frac{7}{2}} }   $  \\
% \end{center}
% \caption{The order of orbits used in the calculations described in
% the text. }\label{Order}
% \end{figure}

\begin{figure}
\epsfig{file=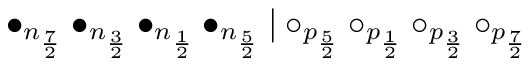}
\caption{The order of orbits used in the calculations described in
the text. }\label{Order}
\end{figure}

\begin{flushleft}
\it {Warmup Phase(The first enlargement stage):}
\end{flushleft}
This phase of the DMRG procedure is aimed at defining a good
initial set of states to use in each block. In our calculations,
we follow the procedure schematically illustrated in Figure 3,
whereby we begin by adding two orbits at one end of the chain of
sites, followed by a subsequent enlargement at the other end,
followed by gradual further enlargements from the two ends of the
chain. In this initialization phase, the last truncated block on
the opposite end of the chain is used as the medium. As should be
clear, the Superblock so obtained does not contain all of the
orbits. Hence the Superblock must be diagonalized for all relevant
neutron and proton numbers in the $J=0^+$ space, to permit some
particles to reside in orbits not included. A mixed density matrix
is constructed from the corresponding eigenstates which is then
used to truncate the system to the $m$ most important states for
each block of orbits. This process is repeated until the middle of
the chain, at which point all neutron and proton orbits have been
treated.

% \begin{figure}
% 
% \begin{tabular}{lcr}
% $B_{r+1} \rightarrow$& &  $ \leftarrow M$\\
% \fbox { \fbox { $\bullet_{\text{n}_{\frac{7}{2}} } %
%                  \bullet_{\text{n}_{\frac{3}{2}} } $} %
%                 $\bullet_{\text{n}_{\frac{1}{2}} }$ }%
% &$ %
%               \bullet_{\text{n}_{\frac{5}{2}}}
%                 \circ_{\text{p}_{\frac{5}{2}} }   %  %
%                 \circ_{\text{p}_{\frac{1}{2}} } $%
% &\fbox { $     %
%                 \circ_{\text{p}_{\frac{3}{2}} }   %
%                 \circ_{\text{p}_{\frac{7}{2}} }   $ } \\
%                 \\
% $M \rightarrow $& &  $\leftarrow B_{r+1}$\\
% \fbox {  $\bullet_{\text{n}_{\frac{7}{2}} } %
%               \bullet_{\text{n}_{\frac{3}{2}} }%
%               \bullet_{\text{n}_{\frac{1}{2}} }$ }%
% &$ %
%               \bullet_{\text{n}_{\frac{5}{2}}}%
%                 \circ_{\text{p}_{\frac{5}{2}} }   %
%                $ %
% &\fbox {  %
%                 $\circ_{\text{p}_{\frac{1}{2}} }$\fbox { $ %
%                 \circ_{\text{p}_{\frac{3}{2}} }   %
%                 \circ_{\text{p}_{\frac{7}{2}} }   $ } } \\
% 
% \end{tabular}
% \caption{Two consecutive enlargement steps during the warmup phase
% of the DMRG calculations described in the text. In the first, the
% third neutron orbit from the left is added; in the next, the third
% neutron orbit from the right is subsequently
% added.}\label{fig:Warmup}
% \end{figure}

\begin{figure}
\epsfxsize=.9\columnwidth\epsfbox{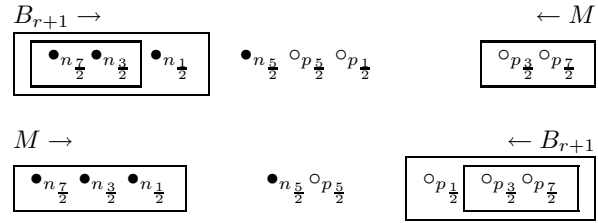}
\caption{Two consecutive enlargement steps during the warmup phase
of the DMRG calculations described in the text. In the first, the
third neutron orbit from the left is added; in the next, the third
neutron orbit from the right is subsequently
added.}\label{fig:Warmup}
\end{figure}

At this point we have an initial choice for the $m$ most important
states for each relevant block.

\begin{flushleft}
{\it The Sweep Phase( Successive enlargement stages):}
\end{flushleft}

We now wish to iteratively update information on each block, by
gradually sweeping through the orbits and using information
initially from the warmup stage and later from the previous sweep
stage to define the medium. We start the sweep phase by adding the
orbits from the center of the chain towards the end. The medium
now consists of one of the like-particle blocks ($M_1$) -
constructed either during the warmup or the previous sweep, and
the unlike-particle block ($M_2$). The fact that our medium
contains two components is why we refer to this as the three-block
algorithm.  The complete growth of the neutron and proton blocks
is repeated in both directions until convergence is attained.
Several consecutive steps in the neutron block enlargement during
a given sweep stage are illustrated schematically in Figure 4.
Because of the great care taken in the warmup initialization, only
three to four sweeps are typically needed for the process to
converge.

% \begin{figure}
% \begin{tabular}{lrr}
% $B_{r+1} \rightarrow$& $\leftarrow M_1$&  $\leftarrow M_2$\\
% \fbox { $\bullet_{\text{n}_{\frac{7}{2}} } %
%               \bullet_{\text{n}_{\frac{3}{2}} } $}%
% &\fbox { $\bullet_{\text{n}_{\frac{1}{2}} } %
%               \bullet_{\text{n}_{\frac{5}{2}}} $ }%
% &\fbox { $  \circ_{\text{p}_{\frac{5}{2}} }   %
%                 \circ_{\text{p}_{\frac{1}{2}} }   %
%                 \circ_{\text{p}_{\frac{3}{2}} }   %
%                 \circ_{\text{p}_{\frac{7}{2}} }   $ } \\
% \fbox { $\bullet_{\text{n}_{\frac{7}{2}} } %
%               \bullet_{\text{n}_{\frac{3}{2}} } %
%               \bullet_{\text{n}_{\frac{1}{2}} } $}%
% &\fbox { $%
%               \bullet_{\text{n}_{\frac{5}{2}}} $ }%
% &\fbox { $  \circ_{\text{p}_{\frac{5}{2}} }   %
%                 \circ_{\text{p}_{\frac{1}{2}} }   %
%                 \circ_{\text{p}_{\frac{3}{2}} }   %
%                 \circ_{\text{p}_{\frac{7}{2}} }   $ } \\
% \fbox { $\bullet_{\text{n}_{\frac{7}{2}} } %
%               \bullet_{\text{n}_{\frac{3}{2}} } %
%               \bullet_{\text{n}_{\frac{1}{2}} } %
%               \bullet_{\text{n}_{\frac{5}{2}} } $}%
% & &\fbox { $  \circ_{\text{p}_{\frac{5}{2}} }   %
%                 \circ_{\text{p}_{\frac{1}{2}} }   %
%                 \circ_{\text{p}_{\frac{3}{2}} }   %
%                 \circ_{\text{p}_{\frac{7}{2}} }   $ } \\
% \end{tabular}
% \caption{Three consecutive steps in the enlargement of the neutron
% block from the left during the sweep phase.}\label{fig:Sweep}
% \end{figure}

\begin{figure}
\epsfxsize=.9\columnwidth\epsfbox{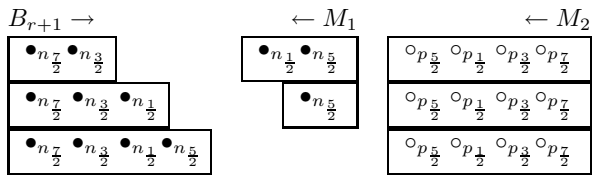}
\caption{Three consecutive steps in the enlargement of the neutron
block from the left during the sweep phase.}\label{fig:Sweep}
\end{figure} 

\begin{flushleft}
{\it Convergence with m:}
\end{flushleft}
The above steps are repeated with increasing $m$ until
satisfactory convergence
in the ground-state energy is obtained.\\

We present here test results for $^{48}$Cr and $^{56}$Ni, using
the three block algorithm described above. We have used the KB3
matrix elements \cite{KB3} used in Refs. \cite{Poves1,Poves2},
with the same set of single-particle energies used there as well.
Excited states for $^{56}$Ni were not included in the analysis
since they are not available for the hamiltonian that we used.

\begin{table}[h]
\caption{Calculated ground-state energies in $MeV$ as a function
of $m$ for $^{48}$Cr. The maximum dimension encountered in the
sweep process is also given. }
\begin{tabular}{|c|c|c|}
\hline
  $m$  &$ E_{GS}$ &   $Max~Dim$\\
\hline
40  & -32.698  &  1,985 \\
60  & -32.763  &  2,859 \\
80  & -32.788  &  3,765 \\
100 & -32.817  &  4,494  \\
120 & -32.840  &  6,367  \\
140 & -32.890  &  8,217  \\
160 & -32.902  &  9,978  \\
180 & -32.944  & 11,062  \\
200 & -32.947  & 12,076  \\
\hline
Exact & -32.953 & 41,355 \\
\hline
\end{tabular}
\end{table}

\begin{table}[h]
\caption{Calculated energies in $MeV$ for low-lying excited states
in $^{48}$Cr, with dimensions shown in brackets.}
\begin{tabular}{|c|c|c|c|c|}
\hline
$m$   & $E_{2^+_1}$ & $E_{4^+_1}$ & $E_{6^+_1}$ & $E_{0^+_2}$ \\
    & ($Dim$) & ($Dim$) & ($Dim$) & ($Dim$) \\
\hline
100 & -31.98    & -30.90    & -29.16      &   -27.97  \\
    & (21,003)  & (33,261)  & (38,652)    &   (44,94) \\
\hline
120 & -32.01    & -30.93    & -29.20      &   -28.06  \\
    &  (28,677) & (42,234)  &  (45,054)   &   (6,367) \\
\hline
140 & -32.04    & -30.98    & -29.26      &   -28.15  \\
    &  (36,706) & (52,254)  &  (52,950)   &   (8,217) \\
\hline
160 & -32.10    & -31.04    & -29.34      &   -28.29   \\
    &  (44,618) & (63,487)  &  (63,537)   &   (9,978)  \\
\hline
180 & -32.13    & -31.09    & -29.43      &   -28.47   \\
    &  (50,030) & (72,616)  &  (74,346)   &   (11,062) \\
\hline
200 & -32.13    & -31.10    & -29.47      &   -28.48   \\
    &  (54,891) & (81,249)  &  (85,168)   &   (12,076) \\
\hline
Exact & -32.15    & -31.13     & -29.55   &   -28.56   \\
      & (182,421) &  (246,979) &  (226,259) & (41,355)  \\
\hline
\end{tabular}
\end{table}

The results for the ground-state energy of $^{48}$Cr are shown in
Table I. The full shell-model space of $^{48}$Cr consists of
1,963,461 states of which 41,355 have $J^{\pi}=0^+$.

The exact result for the ground state is -32.953 $MeV$. Though the
JDMRG results converge monotonically with $m$ to this value, we need a
substantial portion (about 25\%) of the full space to achieve
agreement to within a few $keV$ of the exact energy.

It should be noted that the results reported in Ref. \cite{Pit06}
were for a different order of single-particle levels, explaining
why the ground-state energies and maximum
dimensions listed in that paper are different from those reported here for the same values of $m$. \\

The results for low-lying excited states are provided in Table 2.
To obtain these results, we construct the relevant hamiltonian
matrices using the optimum block structures obtained at the point
of the lowest ground state energy. Note that we do not target the
excited states themselves in constructing the reduced density
matrix used for the truncation. As a result, the convergence is
not as rapid as for the ground state, but nevertheless quite
acceptable.

\begin{table}[h]
\caption{Calculated ground-state energies in $MeV$ as a function
of $m$ for $^{56}$Ni. The maximum dimensions encountered in the
sweep process are also given.}
\begin{tabular}{|c|c|c|}
\hline
$m$  & $E_{GS}$ & $Max~Dim$  \\
\hline
80    &  -78.351 & 72,023    \\
100   &  -78.363 &  83,773   \\
120   & -78.372 & 102,690   \\
140   & -78.376 & 119,797    \\
160   & -78.390 & 136,073    \\
180   & -78.393 & 162,019    \\
200   & -78.399 & 192,878   \\
\hline
Exact &-78.46 & 15,443,684   \\
\hline
\end{tabular}
\end{table}

The results for $^{56}$Ni ground state are tabulated in Table 3.
The size of the exact space in an angular momentum basis is
15,443,684. We have carried out these calculations with the same
order of single-particle orbits as for $^{48}$Cr. The fraction of
space required for meaningful convergence is dramatically reduced
for $^{56}$Ni compared to $^{48}$Cr. With about 1\% of the space
we are able to get within around 60 $keV$ of the exact results
\cite{Poves2}.

Earlier calculations for $^{56}$Ni using the DMRG by Papenbrock
and Dean \cite{Pap03}, although done in the m-scheme, were unable
to obtain the ground state energy to better than 400 $keV$.

An interesting feature of our results for both $^{48}Cr$ and
$^{56}Ni$ is the absence of an exponential falloff in the
converged ground state energy as a function of the number of
states retained. Since exponential behavior is often used to
extrapolate to the actual ground state energy, it is worth
commenting on why it does not occur here.

A key difference between our JDMRG algorithm and the usual DMRG is
that our lattice sites vary significantly in size, as they
represent complete single-particle orbits. Often, when an orbit is
added, no truncation is required at that step. When the increment
in $m$ reaches a large enough value that truncation is avoided, we
find a sharp drop in the ground state energy for that $m$ value.
For lower values, truncation always arises at that step in the
sweep and the falloff is less severe. As a result, the energy
tends to fall off in spurts as a function of $m$ rather than as a
smooth exponential, with larger orbits producing a more jagged
behavior. If, however, we carry out our calculations for a wide
enough range of $m$ values, a fit can still be carried out in the
presence of such behavior. Later, we discuss another way of
enhancing exponential falloff in the JDMRG.

In this work, we have described our efforts aimed at developing
the DMRG method as a dynamical truncation strategy for large scale
shell model calculations of atomic nuclei. Following a brief
description of the usual DMRG procedure, we discussed a specific
three-block angular-momentum-conserving algorithm for carrying out
such calculations. We presented results for two nuclei $^{48}$Cr
and $^{56}$Ni. Although a large fraction of the space was required
to achieve results of high accuracy for $^{48}$Cr, the fraction of
the space required for $^{56}$Ni was much smaller. This bodes well
for the future usefulness of the method for even larger problems.

There are several issues that we plan to explore in the near
future. On the one hand, we would like to extend our test analysis
to a broader range of nuclei, to more meaningfully explore the
requisite fraction of the space required for convergence. We also
plan to redo our calculations for $^{56}$Ni to use the GXPF1A
interaction \cite{GXPF1A}, for which exact results are available
not only for the ground state but for the states in the first
deformed band as well \cite{Bro06}. This will help us to better
assess the ability of the method to accurately treat excited
states as well, even when only the ground state is being targeted
in the iterative truncation process.

Finally, to be able to study even heavier nuclei, it will be
necessary to include orbits with even larger angular momenta. For
example, in nuclei beyond $^{56}$Ni the $1g_{9/2}$ orbit becomes
increasingly more important. The larger the orbit the greater is
the computational strain it imposes on the iterative growth
process of the DMRG. A possible way to get around this is to split
such orbits into two or more parts, while still maintaining exact
angular momentum conservation throughout. We plan to test some
possible approaches for splitting such orbits in the context of
the smaller $1f_{7/2}$ orbital and to compare the results that
emerge with those we have already obtained. Hopefully this will
enable us to then turn to even larger orbits and to continue our
applications up in nuclear mass. Use of smaller lattice sites
may also facilitate a more rapid transition to exponential
behavior in the ground state energy.

This work was supported by US National Science Foundation under
grant \# PHY 0553127. We are grateful for the resources provided 
to us by the National Energy Research Scientific Computing Center.
We acknowledge with deep appreciation the
important contributions of Jorge Dukelsky to this project. We also
wish to thank Alfredo Poves for providing the KB3 matrix elements
used in this work.

\end{document}